\documentclass[12pt]{article}
\usepackage{amssymb}
\usepackage{amsmath}
\def\d{\delta}

\def\la{\lambda}

\def\be{\begin{equation}}
\def\ee{\end{equation}}
\def\arr{\begin{array}{rll}}
\def\ea{\end{array}}
\def\bea{\begin{eqnarray}}
\def\eea{\end{eqnarray}}

\def\N2{$N{=}2$}

\def\>{\rangle}
\def\<{\langle}
\def\+{\dagger}
\def\={\ =\ }

\begin{document}
\renewcommand{\thefootnote}{\arabic{footnote}}
\begin{titlepage}
\setcounter{page}{0}
\begin{flushright}
$\quad$
\end{flushright}
\vskip 1cm
\begin{center}
{\LARGE\bf Higher-derivative mechanics with $\mathcal{N}=2$}\\
\vskip 0.5cm
{\LARGE\bf $l$-conformal Galilei supersymmetry}\\
\vskip 1cm
$
\textrm{\Large Ivan Masterov\ }
$
\vskip 0.7cm
{\it
Laboratory of Mathematical Physics, Tomsk Polytechnic University, \\
634050 Tomsk, Lenin Ave. 30, Russian Federation}
\vskip 0.7cm
{E-mail: masterov@tpu.ru}

\end{center}
\vskip 1cm
\begin{abstract} \noindent
The analysis previously developed in [J. Math. Phys. 55 (2014) 102901] is used to construct systems which hold invariant under $\mathcal{N}=2$ $l$-conformal Galilei superalgebra.  The models describe two different supersymmetric extensions of a free higher-derivative particle. Their Newton-Hooke counterparts are derived by applying appropriate coordinate transformations.
\end{abstract}


\vskip 0.5cm


\end{titlepage}

\noindent
{\bf 1. Introduction}
\vskip 0.5cm

It is well-known that the Galilei algebra can be extended by the generators of dilatations and special conformal transformations in different ways. In general, the set of such extensions is called the $l$-conformal Galilei algebra \cite{Henkel}-\cite{Negro_2}, where $l$ is a positive integer or half-integer parameter. The so-called Schr\"{o}dinger algebra \cite{Niederer_1} and the conformal Galilei algebra \cite{Havas,Lukierski_1} are the most popular instances which correspond to $l=1/2$ and $l=1$, respectively (see \cite{Lukierski_1}-\cite{Colgain}  and references therein).

In recent years, the $l$-conformal Galilei algebra for $l>1$ has attracted much attention \cite{Duval_2}, \cite{Galajinsky_3}-\cite{Andrzejewski_1}. This line of research includes the construction of dynamical realizations \cite{Gomis}-\cite{Galajinsky_1}, \cite{Aizawa_3,PU}, \cite{Masterov_2}-\cite{Andrzejewski_1}, the investigation of supersymmetric extensions \cite{Masterov,Aizawa_1,Aizawa_2,Masterov_2}, the analysis of central charges \cite{Galajinsky_3,Aizawa_1,Aizawa_2,Hosseiny}. Infinite-dimensional generalizations \cite{Galajinsky_3,Masterov,Hosseiny,Masterov_2}, irreducible representations \cite{Aizawa_6,Aizawa_4,Lu,Aizawa_5}, and twist deformations \cite{Dasz} have also been studied. To a large extent, this activity is motivated by the current study of the nonrelativistic version of the AdS/CFT-correspondence \cite{Son,McGreevy}.

If one changes the basis in the $l$-conformal Galilei algebra
\bea\label{change1}
K_{-1}\rightarrow K_{-1}\pm\frac{1}{R^2}K_{1},
\eea
where $K_{-1}$ is the generator of time translations, $K_{1}$ is the generator of special conformal transformations, $\Lambda=\mp\frac{1}{R^{2}}$ is the nonrelativistic cosmological constant, one obtains the structure relations of the so-called $l$-conformal Newton-Hooke algebra \cite{Negro_1,Negro_2,Galajinsky_3}. This algebra is suitable for physical applications in Newton-Hooke spacetime, i.e. in nonrelativistic spacetime with cosmological attraction or repulsion \cite{Gibbons}. Because the $l$-conformal Galilei algebra and the $l$-conformal Newton-Hooke algebra are isomorphic, one can speak about realizations of one and the same algebra in flat spacetime and in nonrelativistic spacetime with cosmological constant, respectively.

So far supersymmetric extensions of the $l$-conformal Galilei algebra have been investigated mostly for $l=1/2$ and $l=1$. It was shown that the Schr\"{o}dinger supersymmetries manifest itself in nonrelativistic spin-$1/2$ particle \cite{Gomis_1}, the nonrelativistic Chern-Simons matter system \cite{Leblanc} and many-body mechanics \cite{Galajinsky_7,Galajinsky_6,Galajinsky_5,Galajinsky_4}. The systematic study of Schr\"{o}dinger superalgebras was given in \cite{Duval_3}. A relation with relativistic superconformal algebras and infinite-dimensional generalizations were studied in \cite{SY_1,SY_2,Sorba,Colgain} and \cite{Henkel_3}, respectively. Various supersymmetric extensions of the conformal Galilei algebra were obtained in \cite{Azcarraga}, \cite{Bagchi}-\cite{S}, \cite{Fedoruk_2} by applying appropriate contraction procedures.

For arbitrary value of the parameter $l$, the $\,\mathcal{N}=1$ and $\,\mathcal{N}=2$ supersymmetric extensions of the $l$-conformal Galilei algebra were formulated in recent works  \cite{Masterov}, \cite{Aizawa_1}, \cite{Aizawa_2}, \cite{Masterov_2}. In particular, in \cite{Masterov_2} dynamical systems which hold invariant under the $\,\mathcal{N}=1$ $l$-conformal Galilei superalgebra were constructed. Yet, the previous studies of
$d=1$, $\mathcal{N}=4$ superconformal mechanics \cite{Wyllard}-\cite{Krivonos_1} indicate that the instance of $\mathcal{N}=2$ is likely to be the maximal number of supersymmetries compatible with translation invariance
of an interacting system. By this reason, it is natural to dwell on the dynamical realizations of the $\,\mathcal{N}=2$ $l$-conformal Galilei superalgebra. The purpose of this work is thus to generalize the results obtained in our recent work \cite{Masterov_2} to the case of $\,\mathcal{N}=2$.

We begin in Section 2 with a brief review of the $l$-conformal Galilei algebra and its $\,\mathcal{N}=2$ supersymmetric extension and then construct dynamical realizations of such a superalgebra in a flat superspace. Associated mechanical systems in Newton-Hooke superspace are considered in Section 3. In Section 4, we summarize our results and discuss possible further developments. Some technical details regarding the derivation of an invariant action functional in Section 2 are presented in Appendix.

\vskip 0.5cm
\noindent
{\bf 2. Higher-derivative mechanics in flat superspace}
\vskip 0.5cm

The $l$-conformal Galilei algebra involves the generators $K_{-1}$ and $K_{1}$ mentioned above, the generator of dilatations $K_{0}$, the set of vector generators $C_i^{(n)}$ with $n=0,1,..,2l$, and the generators of space rotations $M_{ij}$. The non-vanishing structure relations in the algebra read
\begin{align}\label{algebraG}
&
[K_{p},K_{m}]=(m-p)K_{p+m}, &&  [K_{m},C^{(n)}_i]=(n-l(m+1))C_i^{(n+m)},
\nonumber\\[2pt]
&
[M_{ij},C^{(n)}_k]=\delta_{ik} C^{(n)}_j-\delta_{jk} C^{(n)}_i, && [M_{ij},M_{ks}]=\delta_{ik} M_{js}+\delta_{js} M_{ik}-
\delta_{is} M_{jk}-\delta_{jk} M_{is}.
\end{align}

There are two types of $\mathcal{N}=2$ supersymmetric extensions of the $l$-conformal Galilei algebra \cite{Masterov,Aizawa_2} which are called chiral and real. Apart from the generators considered above, both the superalgebras involve a pair of supersymmetry generators $G_{-\frac{1}{2}}$, $\bar{G}_{-\frac{1}{2}}$, the superconformal generators $G_{\frac{1}{2}}$, $\bar{G}_{\frac{1}{2}}$, the generator of $U(1)$ $R$-symmetry transformations $J$, the fermionic partners of the vector generators $L_i^{(n)}$, $\bar{L}_i^{(n)}$ with $n=0,1,..,2l-1$, and the additional bosonic generators $P_i^{(n)}$ with $n=0,1,..,2l-2\gamma$, where
\bea\label{gamma}
\gamma=\left\{
\begin{aligned}
0,&\qquad\mbox{for chiral superalgebra};\\
1,&\qquad\mbox{for real superalgebra}.\\
\end{aligned}
\right.
\eea
It should be noted that for the real $\,\mathcal{N}=2$ supersymmetric extension of the Schr\"{o}dinger algebra, the bosonic generators $P_i^{(n)}$ are absent.

In addition to (\ref{algebraG}), the nonvanishing (anti)commutation relations of the $\,\mathcal{N}=2$ $l$-conformal Galilei superalgebra include
\bea\label{N=2}
&&
\{G_{r},\bar{G}_{s}\}=2iK_{r+s}+(-1)^{r+1/2}\d_{r+s,0}\,J,\qquad\;\, [J,C_i^{(n)}]=\;2l(1-\gamma)\, P_i^{(n)},
\nonumber
\\[5pt]
&&
[G_{r}, C_i^{(n)}]=(n-2l(r+1/2))L_{i}^{(n+r-1/2)},\quad\;\;\, [J,L_i^{(n)}]=\;\;\,i (1+2l(1-\gamma))\, L_i^{(n)},
\nonumber
\\[5pt]
&&
[\bar{G}_{r}, C_i^{(n)}]=(n-2l(r+1/2))\bar{L}_{i}^{(n+r-1/2)},\quad\;\;\, [J,\bar{L}_i^{(n)}]=-i(1+2l(1-\gamma))\,\bar{L}_i^{(n)},
\nonumber
\\[5pt]
&&
[K_{m},P^{(n)}_i]=(n-(l-\gamma)(m+1))P_i^{(n+m)},\quad\; [J,P_i^{(n)}]=-2l(1-\gamma)\,C_i^{(n)},
\nonumber
\\[5pt]
&&
[K_m,L_{i}^{(n)}]=(n-(l-1/2)(m+1))L_{i}^{(n+m)},\;\; [K_m,G_{r}]=(r-m/2)G_{m+r},
\nonumber
\\[5pt]
&&
[K_m,\bar{L}_{i}^{(n)}]=(n-(l-1/2)(m+1))\bar{L}_{i}^{(n+m)},\;\; [K_m,\bar{G}_{r}]=(r-m/2)\bar{G}_{m+r},
\nonumber
\\[5pt]
&&
[G_{r},P_i^{(n)}]=\;\;\, i\Bigl(1+(1-\gamma)[n-2l(r+1/2)-1]\Bigr)L_{i}^{(n+r+\gamma-1/2)},\;[J,G_{r}]=\;\;\,i\, G_{r},
\\[5pt]
&&
[\bar{G}_{r},P_i^{(n)}]=-i\Bigl(1+(1-\gamma)[n-2l(r+1/2)-1]\Bigr)\bar{L}_{i}^{(n+r+\gamma-1/2)},\;[J,\bar{G}_{r}]=-i\, \bar{G}_{r},
\nonumber
\\[5pt]
&&
\{G_{r},\bar{L}_{i}^{(n)}\}=i C_i^{(n+r+1/2)}-\Bigl(1+\gamma[n-(2l-1)(r+1/2)-1]\Bigr) P^{(n-\gamma+r+1/2)}_{i},
\nonumber
\eea
\bea
&&
\{\bar{G}_{r},L_{i}^{(n)}\}=i C_i^{(n+r+1/2)}+\Bigl(1+\gamma[n-(2l-1)(r+1/2)-1]\Bigr) P^{(n-\gamma+r+1/2)}_{i},
\nonumber
\\[5pt]
&&
[M_{ij},A^{(n)}_k]=\delta_{ik} A^{(n)}_j-\delta_{jk} A^{(n)}_i,\qquad A_i^{(n)}=L_i^{(n)},\;\bar{L}_i^{(n)},\; P_i^{(n)}.
\nonumber
\eea

The superalgebra (\ref{algebraG}), (\ref{N=2}) admits a central extension. According to the results in \cite{Galajinsky_3,Aizawa_1,Aizawa_2}, (anti)commutators between the vector generators can be modified as follows:
\bea\label{CE}
&&
[C_i^{(n)},C_j^{(m)}]=(-1)^{n}n!m!\d_{n+m,2l}\lambda_{ij}M,\quad\{L_i^{(n)},\bar{L}_j^{(m)}\}=i(-1)^{n}n!m!\d_{n+m,2l-1}\lambda_{ij}M,
\\[2pt]
&&
\qquad\qquad\qquad\qquad[P_i^{(n)},P_j^{(m)}]=(-1)^{n+\gamma}n!m!\d_{n+m,2l-2\gamma}\lambda_{ij}M,
\nonumber
\eea
where $M$ is a central charge and
\bea
\lambda_{ij}=\left\{
\begin{aligned}
\d_{ij},&\qquad i,j=1,2,..,d,&\qquad\mbox{for half-integer $l$};\\
\epsilon_{ij},&\qquad i,j=1,2,&\qquad\mbox{for integer $l$},\\
\end{aligned}
\right.
\eea
with $\epsilon_{12}=1$.

The central extension (\ref{CE}) allows one to obtain a realization of all the scalar generators as well as the generators of space rotations in terms of the quadratic combinations of the vector generators\footnote{Throughout the work the summation over repeated spatial indices is understood.} \cite{Aizawa_1,Aizawa_2}
\bea
\begin{aligned}
&
K_{n}=\la_{ij}(\sum_{k=0}^{2l}\sigma_{k,0}^{n}C^{(2l-k)}_{i}C^{(k+n)}_{j}
+2i\sum_{k=0}^{2l-1}\sigma_{k,1}^{n}L^{(2l-k-1)}_{i}\bar{L}^{(k+n)}_{j}+(-1)^{\gamma}\sum_{k=0}^{2l-2\gamma}\sigma_{k,2\gamma}^{n}P^{(2l-2\gamma-k)}_i P^{(k+n)}_{j}),
\\
&
G_{r}=2\la_{ij}\sum_{k=0}^{2l}\sigma_{k,0}^{-1}\bigl(C^{(2l-k+r+1/2)}_{i}+i(1-\gamma[(2l-1)(r-1/2)+k])P^{(2l-k+r+1/2-\gamma)}_i\bigr)
L^{(k-1)}_j,
\\
&
\bar{G}_{r}=2\la_{ij}\sum_{k=0}^{2l}\sigma_{k,0}^{-1}\bigl(C^{(2l-k+r+1/2)}_{i}-i(1-\gamma[(2l-1)(r-1/2)+k])P^{(2l-k+r+1/2-\gamma)}_{i}\bigr)
\bar{L}^{(k-1)}_{j},
\\
&
J=\frac{2l-2l\gamma+1}{2}\la_{ij}\sum_{k=0}^{2l-1}\nu_{k,1}\left(L^{(2l-k-1)}_{i}\bar{L}^{(k)}_{j}-\bar{L}^{(2l-k-1)}_i L^{(k)}_j\right)+
2l(1-\gamma)\la_{ij}\sum_{k=0}^{2l}\nu_{k,0}C^{(2l-k)}_{i}P^{(k)}_{j},
\\
&
M_{ij}=\sum_{k=0}^{2l}\nu_{k,0}\, C_i^{(2l-k)}C_j^{(k)}+(-1)^{\gamma} \sum_{k=0}^{2l-2\gamma}\nu_{k,2\gamma}P_i^{(2l-2\gamma-k)}P_j^{(k)}+
\\
&
\qquad\qquad\qquad\qquad\qquad\qquad\qquad\qquad\qquad\qquad\qquad+i\sum_{k=0}^{2l-1}\nu_{k,1}(L_{i}^{(2l-k-1)} \bar{L}_{j}^{(k)}+\bar{L}_{i}^{(2l-k-1)} L_{j}^{(k)}).
\end{aligned}
\nonumber
\eea
\bea\label{BP}
\quad
\eea
For integer $l$ the generator of spatial rotations has the form
\bea\label{BP1}
M_{12}=-\sum_{k=0}^{2l}\frac{\nu_{k,0}}{2}\, C_i^{(2l-k)}C_i^{(k)}-i\sum_{k=0}^{2l-1}\nu_{k,1}L_{i}^{(2l-k-1)} \bar{L}_{i}^{(k)}-(-1)^{\gamma} \sum_{k=0}^{2l-2\gamma}\frac{\nu_{k,2\gamma}}{2}P_i^{(2l-2\gamma-k)}P_i^{(k)},
\eea
where we denoted
\bea
\sigma_{k,s}^{n}=\frac{1}{2}\left(k-(l-s/2)(n+1)\right)\nu_{k,s},\qquad \nu_{k,s}=\frac{(-1)^{2l-k-s}}{M k!(2l-k-s)!}.
\nonumber
\eea

In order to construct a dynamical realization of the superalgebra (\ref{algebraG}), (\ref{N=2}), (\ref{CE}), we adopt the strategy suggested in \cite{Masterov_2}. As the first step, we construct a system which possesses the vector integrals of motion satisfying the relations (\ref{CE}) under some graded Poisson bracket. This can be done by applying the conventional method of nonlinear realizations \cite{Coleman_1,Coleman_2} to the subalgebra which is formed by the generator of time translations together with all the vector generators and the central charge $M$. This results in the action functional
\bea\label{action}
S=\frac{1}{2}\int\,dt\,\lambda_{ij}\,\left(x_i\frac{d^{2l+1}x_j}{dt^{2l+1}}-i\psi_i\frac{d^{2l}\bar{\psi}_j}{dt^{2l}}- i\bar{\psi}_i\frac{d^{2l}\psi_j}{dt^{2l}}+(-1)^{\gamma}\,z_i\frac{d^{2l-2\gamma+1}z_j}{dt^{2l-2\gamma+1}}\right).
\eea
This action describes a free $\,\mathcal{N}=2$ higher-derivative superparticle whose configuration space of degrees of freedom involves the bosonic coordinates $x_i$, the fermionic coordinates $\psi_i$ and $\bar{\psi}_i$, and extra bosonic coordinates $z_i$. The bosonic limit of (\ref{action}) has been recently studied in \cite{Gomis}-\cite{Gonera_5}, \cite{Gonera_3}-\cite{Gonera_2}.  Some technical details regarding the derivation of this
action are gathered in Appendix.

The action (\ref{action}) is invariant under the transformations
\bea\label{transf}
\d x_i=\sum_{n=0}^{2l}a_i^{(n)}t^n,\quad \d \bar{\psi}_i=\sum_{n=0}^{2l-1}\bar{\xi}_i^{(n)}t^n,\quad \d \psi_i=\sum_{n=0}^{2l-1}\xi_i^{(n)}t^n,\quad \d z_i=\sum_{n=0}^{2l-2\gamma}b_i^{(n)}t^n,
\eea
which yield the following integrals of motion:
\bea\label{INT}
\begin{aligned}
&
C_i^{(n)}=\la_{ij}\sum_{k=0}^{n}\frac{(-1)^{k+1}n!}{(n-k)!}t^{n-k} x_j^{(2l-k)},\quad\;\, L_i^{(n)}=i\la_{ij}\sum_{k=0}^{n}\frac{(-1)^{k}n!}{(n-k)!}t^{n-k} \psi_j^{(2l-k-1)},
\\
&
\bar{L}_i^{(n)}=i\la_{ij}\sum_{k=0}^{n}\frac{(-1)^{k}n!}{(n-k)!}t^{n-k} \bar{\psi}_j^{(2l-k-1)},\quad P_i^{(n)}=\la_{ij}\sum_{k=0}^{n}\frac{(-1)^{k+\gamma+1}n!}{(n-k)!}t^{n-k} z_j^{(2l-2\gamma-k)},
\end{aligned}
\eea
where $f^{(n)}(t)\equiv\frac{d^n f(t)}{dt^n}$, $f^{(0)}(t)\equiv f(t)$.

At the next step, let us introduce the graded Poisson bracket
\bea\label{PB}
\begin{aligned}
&
[A,B\}=\la_{ij}\left(\sum_{n=0}^{2l}(-1)^n \frac{\partial A}{\partial x_i^{(n)}}\frac{\partial B}{\partial x_j^{(2l-n)}}-i\sum_{n=0}^{2l-1}(-1)^{n}\frac{\overleftarrow{\partial}A}{\partial\psi_i^{(n)}}\frac{\overrightarrow{\partial}B}{\partial\bar{\psi}_j^{(2l-n-1)}}-\right.
\\[2pt]
&
\qquad\qquad\qquad\left.-i\sum_{n=0}^{2l-1}(-1)^{n}\frac{\overleftarrow{\partial}A}{\partial\bar{\psi}_i^{(n)}}\frac{\overrightarrow{\partial}B}{\partial\psi_j^{(2l-n-1)}}+\sum_{n=0}^{2l-2\gamma}(-1)^{n+\gamma} \frac{\partial A}{\partial z_i^{(n)}}\frac{\partial B}{\partial z_j^{(2l-2\gamma-n)}}\right).
\end{aligned}
\eea
This can be viewed as an $\mathcal{N}=2$ generalization of the bracket considered recently in \cite{Masterov_2}. It is straightforward to verify that the vector integrals of motion (\ref{INT}) obey the structure relations (\ref{CE}) under the graded bracket (\ref{PB}) with $M=1$. Therefore, the expressions (\ref{BP}) and (\ref{BP1}) provide further integrals of motion  which together with (\ref{INT}) form a representation of the $\mathcal{N}=2$ $l$-conformal Galilei superalgebra (\ref{algebraG}), (\ref{N=2}), (\ref{CE}). Symmetry transformations of the action functional (\ref{action}) which correspond to the integrals (\ref{BP}), (\ref{BP1}) read
\bea\label{trN2}
\begin{aligned}
&
K_{n}:\qquad\;\, \d t=t^{n+1} a_{n},\qquad \d x_i=l(n+1) t^{n} a_n x_i,\qquad \d\psi_i=(l-1/2)(n+1) t^n a_n \psi_i,
\\[5pt]
&
\,\qquad\qquad\quad\qquad\d\bar{\psi}_i=(l-1/2)(n+1) t^n a_n \bar{\psi}_i,\qquad \d z_i=(l-\gamma)(n+1) t^{n} a_n z_i;
\\[10pt]
&
G_{-\frac{1}{2}}:\qquad\quad \d x_i=i\psi_i\alpha_{-\frac{1}{2}},\qquad\;\d z_i=-\psi_i^{(\gamma)}\alpha_{-\frac{1}{2}},\qquad\; \d\bar{\psi}_i=(x_i^{(1)}+iz_i^{(1-\gamma)})\alpha_{-\frac{1}{2}};
\\[2pt]
&
\bar{G}_{-\frac{1}{2}}:\qquad\quad \d x_i=i\bar{\psi}_i\bar{\alpha}_{-\frac{1}{2}},\qquad\;\d z_i=\bar{\psi}_i^{(\gamma)}\bar{\alpha}_{-\frac{1}{2}},\qquad\quad \d\psi_i=(x_i^{(1)}-iz_i^{(1-\gamma)})\bar{\alpha}_{-\frac{1}{2}};
\\[10pt]
&
G_{\frac{1}{2}}:\qquad\qquad\qquad \d x_i=it\psi_i\,\alpha_{\frac{1}{2}},\;\qquad\qquad \d z_i=(-t\psi_i^{(\gamma)}+(2l-1)\gamma\psi_i)\alpha_{\frac{1}{2}},
\\[2pt]
&
\qquad\qquad\qquad\qquad\qquad\d\bar{\psi}_i=(tx_i^{(1)}-2lx_i+itz_i^{(1-\gamma)}-2il(1-\gamma)z_i)\alpha_{\frac{1}{2}};
\\[2pt]
&
\bar{G}_{\frac{1}{2}}:\qquad\qquad\qquad\, \d x_i=it\bar{\psi}_i\,\bar{\alpha}_{\frac{1}{2}},\;\qquad\qquad \d z_i=(t\bar{\psi}_i^{(\gamma)}-(2l-1)\gamma\bar{\psi}_i)\bar{\alpha}_{\frac{1}{2}},
\\[2pt]
&
\qquad\qquad\qquad\qquad\qquad\d\psi_i=(tx_i^{(1)}-2lx_i-itz_i^{(1-\gamma)}+2il(1-\gamma)z_i)\bar{\alpha}_{\frac{1}{2}};
\\[10pt]
&
J:\qquad\qquad\qquad\;\,\;\d x_i=2l(1-\gamma)\nu z_i,\qquad\qquad\d\psi_i=i(2l-2l\gamma+1)\nu\psi_i,
\\[5pt]
&
\qquad\qquad\quad\qquad\quad\d z_i=-2l(1-\gamma)\nu x_i,\qquad\quad\;\d\bar{\psi}_i=-i(2l-2l\gamma+1)\nu\bar{\psi}_i,
\\[10pt]
&
M_{ij}:\qquad\d x_i=w_{ij}x_j,\quad \d\psi_i=w_{ij}\psi_j,\quad\, \d\bar{\psi}_i=w_{ij}\bar{\psi}_j,\quad\, \d z_i=w_{ij}z_j,\quad\, (w_{ij}=-w_{ji}).
\end{aligned}
\eea

Thus, a free higher-derivative superparticle (\ref{action}) provides a dynamical realization of the $\,\mathcal{N}=2$ $l$-conformal Galilei superalgebra (\ref{algebraG}), (\ref{N=2}), (\ref{CE}).
It should be remembered that the bosonic generators $P_i^{(n)}$ and the coordinates $z_i^{(n)}$ must be discarded for the case of the real $\,\mathcal{N}=2$ Schr\"{o}dinger superalgebra.

\vskip 0.5cm
\noindent
{\bf 3. Associated mechanics in Newton-Hooke superspace}
\vskip 0.5cm
Inspired by the results in \cite{Masterov_2}, let us introduce an analogue of Niederer's coordinate transformations \cite{Niederer} which link dynamical realizations of $\,\mathcal{N}=2$ $l$-conformal Galilei superalgebra in flat superspace to those in Newton-Hooke superspace. For the case of a negative cosmological constant we set
\bea\label{Nied}
&&
t'=R\tan(t/R),\qquad x_i'(t')=x_i(t)/\cos^{2l}(t/R),\qquad \psi_i'(t')=\psi_i(t)/\cos^{2l-1}(t/R),
\nonumber
\\[7pt]
&&
\qquad\qquad\bar{\psi}_i'(t')=\bar{\psi}_i(t)/\cos^{2l-1}(t/R),\qquad z_i'(t')=z_i(t)/\cos^{2l-2\gamma}(t/R).
\eea
For half-integer $l$ implementation of (\ref{Nied}) in (\ref{action}) results in the action functional
\bea\label{PU1}
\begin{aligned}
&
\;\, S=\frac{1}{2}\int dt\left(x_i\prod_{k=1}^{l+\frac{1}{2}}\left(\frac{d^2}{dt^2}+\frac{(2k-1)^2}{R^2}\right)x_i- i\psi_i\prod_{k=1}^{l-\frac{1}{2}}\left(\frac{d^2}{dt^2}+\frac{(2k)^2}{R^2}\right)\dot{\bar{\psi}}_i-\right.
\\[5pt]
&
\left.\qquad\qquad- i\bar{\psi}_i\prod_{k=1}^{l-\frac{1}{2}}\left(\frac{d^2}{dt^2}+\frac{(2k)^2}{R^2}\right)\dot{\psi}_i +(-1)^{\gamma}\,z_i\prod_{k=1}^{l-\gamma+\frac{1}{2}}\left(\frac{d^2}{dt^2}+\frac{(2k-1)^2}{R^2}\right)z_i\right),
\end{aligned}
\eea
while for integer $l$ one has
\bea\label{PU2}
\begin{aligned}
&
S=\frac{1}{2}\int dt\, \epsilon_{ij}\left(x_i\prod_{k=1}^{l}\left(\frac{d^2}{dt^2}+\frac{(2k)^2}{R^2}\right)\dot{x}_j- i\psi_i\prod_{k=1}^{l}\left(\frac{d^2}{dt^2}+\frac{(2k-1)^2}{R^2}\right)\bar{\psi}_j-\right.
\\[5pt]
&
\left.\qquad\quad\;\;\, - i\bar{\psi}_i\prod_{k=1}^{l}\left(\frac{d^2}{dt^2}+\frac{(2k-1)^2}{R^2}\right)\psi_j+ (-1)^{\gamma}\,z_i\prod_{k=1}^{l-\gamma}\left(\frac{d^2}{dt^2}+\frac{(2k)^2}{R^2}\right)\dot{z}_j\right).
\end{aligned}
\eea
These actions describe $\mathcal{N}=2$ supersymmetric generalizations of the so-called Pais-Uhlenbeck oscillator \cite{Pais} (for a review see also \cite{Smilga}) for a particular choice of its frequencies. Recently, the bosonic limit of the models (\ref{PU1}) and (\ref{PU2}) has been extensively studied in \cite{Galajinsky_1}, \cite{PU}, \cite{Andrzejewski,Andrzejewski_1}.

The coordinate transformations (\ref{Nied}) allow one to readily obtain symmetry transformations and associated integrals of motion from the expressions (\ref{transf}), (\ref{INT}), (\ref{trN2}), and (\ref{BP}). For example, the application of (\ref{Nied}) to (\ref{transf}) yields
\bea\label{transf1}
\begin{aligned}
&
\d x_i=\sum_{n=0}^{2l}a_i^{(n)}R^n \sin^{n}{\frac{t}{R}}\cos^{2l-n}{\frac{t}{R}},\quad\quad
\d\bar{\psi}_i=\sum_{n=0}^{2l-1}\bar{\xi}_i^{(n)}R^n \sin^{n}{\frac{t}{R}}\cos^{2l-n-1}{\frac{t}{R}},
\\[2pt]
&
\d\psi_i=\sum_{n=0}^{2l-1}\xi_i^{(n)}R^n \sin^{n}{\frac{t}{R}}\cos^{2l-n-1}{\frac{t}{R}},\quad
\d z_i=\sum_{n=0}^{2l-2\gamma}b_i^{(n)}R^n \sin^{n}{\frac{t}{R}}\cos^{2l-2\gamma-n}{\frac{t}{R}}.
\end{aligned}
\eea
The associated conserved vector charges obey the structure relations (\ref{CE}) under the graded Poisson bracket which is obtained from (\ref{PB}) with the help of (\ref{Nied}) (see also a related discussion in \cite{Masterov_2}). The graded bracket enables one to automatically produce additional integrals of motion. In particular, taking into account (\ref{change1}) and the following the redefinition:
\bea\label{change2}
G_{-\frac{1}{2}}\,\rightarrow\,G_{-\frac{1}{2}}+\frac{i}{R}G_{\frac{1}{2}},\qquad \bar{G}_{-\frac{1}{2}}\,\rightarrow\,\bar{G}_{-\frac{1}{2}}-\frac{i}{R}\bar{G}_{\frac{1}{2}},
\eea
one finds the infinitesimal symmetry transformations of the form
\bea\label{NHtr}
&&
K_{0}:\qquad\;\,\d t=\frac{R}{2}\sin{\frac{2t}{R}}\,a_{0},\qquad\;
\d x_i=l\cos{\frac{2t}{R}}\,x_i\, a_{0},\qquad\;
\d\psi_i=\Big(l-\frac{1}{2}\Big)\cos{\frac{2t}{R}}\,\psi_i\,a_{0},
\nonumber
\\[2pt]
&&
\qquad\qquad\qquad\quad\d\bar{\psi}_i=\Big(l-\frac{1}{2}\Big)\cos{\frac{2t}{R}}\,\bar{\psi}_i\,a_{0},\qquad\;\d z_i=(l-\gamma)\cos{\frac{2t}{R}}\,z_i\, a_{0};
\nonumber
\\[10pt]
&&
K_{1}:\qquad\;\,\d t=R^2\sin^2{\frac{t}{R}}\,a_{1},\quad\;\;\d x_i=lR\sin{\frac{2t}{R}}\,x_i\,a_{1},\quad\;\;\,
\d\psi_i=\Big(l-\frac{1}{2}\Big)R\sin{\frac{2t}{R}}\psi_i a_{1},
\nonumber
\\[2pt]
&&
\qquad\qquad\qquad\quad\d\bar{\psi}_i=\Big(l-\frac{1}{2}\Big)R\sin{\frac{2t}{R}}\,\bar{\psi}_i\, a_{1},\quad\;\,\d z_i=(l-\gamma)R\sin{\frac{2t}{R}}\,z_i\, a_{1};
\nonumber
\\[7pt]
&&
G_{-\frac{1}{2}}:\qquad \d x_i=i e^{\frac{it}{R}}\,\psi_i\,\alpha_{-\frac{1}{2}},\qquad \d z_i=e^{\frac{it}{R}}\left(-\psi_i^{(\gamma)}+\frac{2l-1}{R} i\gamma\psi_i\right)\alpha_{-\frac{1}{2}},
\nonumber
\\[2pt]
&&
\qquad\qquad\qquad\d\bar{\psi}_i=e^{\frac{it}{R}}\left(x_i^{(1)}-\frac{2il}{R}x_i+iz_i^{(1-\gamma)} +\frac{2l(1-\gamma)}{R}z_i\right)\alpha_{-\frac{1}{2}};
\nonumber
\eea
\bea
&&
\bar{G}_{-\frac{1}{2}}:\qquad \d x_i=i e^{-\frac{it}{R}}\,\bar{\psi}_i\,\bar{\alpha}_{-\frac{1}{2}},\qquad \d z_i=e^{-\frac{it}{R}}\left(\bar{\psi}_i^{(\gamma)}+\frac{2l-1}{R} i\gamma\bar{\psi}_i\right)\bar{\alpha}_{-\frac{1}{2}},
\nonumber
\\[2pt]
&&
\qquad\qquad\qquad\d\psi_i=e^{-\frac{it}{R}}\left(x_i^{(1)}+\frac{2il}{R}x_i-iz_i^{(1-\gamma)} +\frac{2l(1-\gamma)}{R}z_i\right)\bar{\alpha}_{-\frac{1}{2}};
\nonumber
\\[2pt]
&&
G_{\,\frac{1}{2}}:\qquad\;\d x_i=iR\sin{\frac{t}{R}}\,\psi_i\,\alpha_{\frac{1}{2}},\qquad \d z_i=\left(-R\sin{\frac{t}{R}}\psi_i^{(\gamma)}+(2l-1)\gamma\cos{\frac{t}{R}}\psi_i\right)\alpha_{\frac{1}{2}},
\nonumber
\\[2pt]
&&
\qquad\qquad\qquad\d\bar{\psi}_i=\left(R\sin{\frac{t}{R}}x_i^{(1)}-2l\cos{\frac{t}{R}}x_i+ iR\sin{\frac{t}{R}}z_i^{(1-\gamma)}-2il(1-\gamma)\cos{\frac{t}{R}}z_i\right)\alpha_{\frac{1}{2}};
\nonumber
\\[2pt]
&&
\bar{G}_{\,\frac{1}{2}}:\qquad\;\d x_i=iR\sin{\frac{t}{R}}\,\bar{\psi}_i\,\bar{\alpha}_{\frac{1}{2}},\qquad \d z_i=\left(R\sin{\frac{t}{R}}\bar{\psi}_i^{(\gamma)}-(2l-1)\gamma\cos{\frac{t}{R}}\bar{\psi}_i\right)\bar{\alpha}_{\frac{1}{2}},
\nonumber
\\[2pt]
&&
\qquad\qquad\qquad\d\psi_i=\left(R\sin{\frac{t}{R}}x_i^{(1)}-2l\cos{\frac{t}{R}}x_i- iR\sin{\frac{t}{R}}z_i^{(1-\gamma)}+2il(1-\gamma)\cos{\frac{t}{R}}z_i\right)\bar\alpha_{\frac{1}{2}}.
\nonumber
\eea
Transformations associated with $K_{-1}$, $J$, $M_{ij}$ hold the same form as in (\ref{INT}).

For the case of a positive cosmological constant Niederer-like transformations can be obtained from (\ref{Nied}) by a formal change $R\rightarrow iR$. Note that the flat space limit $R\rightarrow\infty$ reproduces the model (\ref{action}).

\vskip 0.5cm
\noindent
{\bf 4. Conclusion}
\vskip 0.5cm

To summarize, in this paper, we have constructed the simplest dynamical realizations of $\mathcal{N}=2$ $l$-conformal Galilei superalgebra in flat superspace and in Newton-Hooke superspace. In the latter case,
the model can be interpreted as an $\mathcal{N}=2$ supersymmetric extension of the Pais-Uhlenbeck oscillator for a particular choice of its frequencies. Coordinate transformations which link the models have been constructed.

In this work, we made the first step towards investigating mechanical systems exhibiting $\mathcal{N}=2$ $l$-conformal Galilei supersymmetry. We hope that further developments
will demonstrate advantages of supersymmetric $l$-conformal theories. The first issue to mention is that the integrals of motion$K_{-1}$, $G_{-\frac{1}{2}}$ and $\bar{G}_{-\frac{1}{2}}$ for the models (\ref{PU1}) and (\ref{PU2}) obey the structure relations
\bea
\{G_{-\frac{1}{2}},\bar{G}_{-\frac{1}{2}}\}=2i\left(K_{-1}-\frac{1}{R}J\right),\quad[K_{-1},G_{-\frac{1}{2}}]=\frac{i}{R}G_{-\frac{1}{2}},\quad [K_{-1},\bar{G}_{-\frac{1}{2}}]=-\frac{i}{R}\bar{G}_{-\frac{1}{2}}.
\nonumber
\eea
It would be interesting to modify the realizations constructed above so as to bring the structure relations to the standard form. This issue will be studied elsewhere.
Another interesting problem to tackle is to construct dynamical realizations of $\,\mathcal{N}=2$ $l$-conformal Galilei superalgebra without higher-derivatives. It is also worth generalizing the results in \cite{Henkel_4}, \cite{Henkel_5} to incorporate $\mathcal{N}=2$ supersymmetry. Investigation of $l$-conformal Galilei superalgebras for $\,\mathcal{N}>2$ and construction of their dynamical realizations may potentially raise many interesting issues. In this context it would be interesting to construct $l$-conformal extensions of the systems obtained recently in \cite{Toppan} and the superparticles with rigidity in \cite{Plyushchay_1,Krivonos_2}.

\newpage
\noindent
{\bf Acknowledgements}
\vskip 0.5cm
We thank A. Galajinsky for the comments on the manuscript and E. \'{O} Colg\'{a}in and K. Andrzejewski for their interest in the work and the useful correspondence. This work was supported
by the Dynasty Foundation, RFBR grant 14-02-31139-Mol, the MSE program "Nauka" under
the project 3.825.2014/K, and the TPU grant LRU.FTI.123.2014.

\vskip 0.5cm
\noindent
{\bf Appendix. Higher-derivative superparticle action from nonlinear realization}
\vskip 0.5cm

Let us apply the method of nonlinear realizations \cite{Coleman_1,Coleman_2} to the subalgebra which is formed by the generator of time translations together with all the vector generators and the central charge $M$. The left multiplication by a subgroup element
\bea
e^{aK_{-1}}e^{a_i^{(n)}C_i^{(n)}}e^{i\xi_i^{(n)}L_i^{(n)}+i\bar{\xi}_i^{(n)}\bar{L}_i^{(n)}}e^{b_i^{(n)}P_i^{(n)}}e^{\phi M}
\nonumber
\eea
determines the action on the space
\bea
G=e^{tK_{-1}}e^{x_i^{(n)}C_i^{(n)}}e^{i\psi_i^{(n)}L_i^{(n)}+i\bar{\psi}_i^{(n)}\bar{L}_i^{(n)}}e^{z_i^{(n)}P_i^{(n)}}e^{\varphi M}
\nonumber
\eea
parameterized by the even $t,\,x_i^{(n)},\,z_i^{(n)},\,\varphi$, and odd $\psi_i^{(n)},\,\bar{\psi}_i^{(n)}$ coordinates. Infinitesimal coordinate transformations which correspond to this action read
$$
\d x_i^{(n)}=\sum_{k=n}^{2l}\frac{(-1)^{n-k}k!}{n!(k-n)!}t^{k-n}a_i^{(k)},\quad\qquad\d \psi_i^{(n)}=\sum_{k=n}^{2l-1}\frac{(-1)^{n-k}k!}{n!(k-n)!}t^{k-n}\xi_i^{(k)},
$$
$$
\d\bar{\psi}_i^{(n)}=\sum_{k=n}^{2l-1}\frac{(-1)^{n-k}k!}{n!(k-n)!}t^{k-n}\bar{\xi}_i^{(k)},\quad\qquad\;\d z_i^{(n)}=\sum_{k=n}^{2l-2\gamma}\frac{(-1)^{n-k}k!}{n!(k-n)!}t^{k-n}b_i^{(k)},\eqno{(A1)}
$$
$$
\d t=a,\qquad\d\varphi=\phi+\sum_{n=0}^{2l}\sum_{k=n}^{2l}\upsilon_{k,0}^{n}\,t^{k-n}a_i^{(k)}\lambda_{ij}x_j^{(2l-n)}+ i\sum_{n=0}^{2l-1}\sum_{k=n}^{2l-1}\upsilon_{k,1}^{n}\,t^{k-n}\xi_i^{(k)}\lambda_{ij}\bar{\psi}_j^{(2l-n-1)}+
$$
$$
+i\sum_{n=0}^{2l-1}\sum_{k=n}^{2l-1}\upsilon_{k,1}^{n}\,t^{k-n}\bar{\xi}_i^{(k)}\lambda_{ij}\psi_j^{(2l-n-1)}+ (-1)^{\gamma}\sum_{n=0}^{2l-2\gamma}\sum_{k=n}^{2l-2\gamma}\upsilon_{k,2\gamma}^{n}\,t^{k-n}b_i^{(k)}\lambda_{ij}z_j^{(2l-2\gamma-n)},
$$
where we denoted
\bea
\upsilon_{k,s}^{n}=(-1)^k\,\frac{k!\,(2l-n-s)!}{2\,(k-n)!}.
\nonumber
\eea
The generators of these transformations form the subalgebra of $\,\mathcal{N}=2$ $l$-conformal Galilei superalgebra (\ref{algebraG}), (\ref{N=2}), (\ref{CE}) which involve $K_{-1}$, $C_i^{(n)}$, $L_i^{(n)}$, $\bar{L}_i^{(n)}$, $P_i^{(n)}$ and $M$.

Then let us construct the left-invariant Maurer-Cartan one-forms
\bea
G^{-1}dG=\omega_K K_{-1}+\omega_{C,i}^{(n)}C_i^{(n)}+i\,\omega_{L,i}^{(n)}L_i^{(n)}+ i\,\omega_{\bar{L},i}^{(n)}\bar{L}_i^{(n)}+\omega_{P,i}^{(n)}P_i^{(n)}+\omega_M M,
\nonumber
\eea
where\footnote{By definition $x_i^{(2l+1)}=\psi_i^{(2l)}=\bar{\psi}_i^{(2l)}=z_i^{(2l-2\gamma+1)}=0$. }
\bea
&&
\omega_{C,i}^{(n)}=dx_i^{(n)}+(n+1)x_i^{(n+1)}dt,\qquad\quad\omega_{L,i}^{(n)}=d\psi_i^{(n)}+(n+1)\psi_i^{(n+1)}dt,
\nonumber
\\[5pt]
&&
\omega_{\bar{L},i}^{(n)}=d\bar{\psi}_i^{(n)}+(n+1)\bar{\psi}_i^{(n+1)}dt,\qquad\quad\omega_{P,i}^{(n)}=dz_i^{(n)}+(n+1)z_i^{(n+1)}dt,
\nonumber
\\[5pt]
&&
\omega_K=dt,\qquad\omega_M=d\varphi+\sum_{n=0}^{2l}\upsilon_{n,0}^{n}\,\omega_{C,i}^{(n)}\,\lambda_{ij}\,x_j^{(2l-n)}+ i\sum_{n=0}^{2l-1}\upsilon_{n,1}^{n}\,\omega_{L,i}^{(n)}\,\lambda_{ij}\,\bar{\psi}_j^{(2l-n-1)}+
\nonumber
\\[2pt]
&&
+i\sum_{n=0}^{2l-1}\upsilon_{n,1}^{n}\,\omega_{\bar{L},i}^{(n)}\,\lambda_{ij}\,\psi_j^{(2l-n-1)}+ (-1)^{\gamma}\sum_{n=0}^{2l-2\gamma}\upsilon_{n,2\gamma}^{n}\,\omega_{P,i}^{(n)}\,\lambda_{ij}\,z_j^{(2l-2\gamma-n)}.
\nonumber
\eea
By construction, these forms are invariant under all the transformations (A1).

Some degrees of freedom can be reduced by setting some of the one-forms to vanish \cite{Ivanov}. If we take $t$ to be a temporal coordinate, then the restrictions
\bea
\omega_{C,i}^{(n)}=0,\qquad \omega_{L,i}^{(n)}=0,\qquad \omega_{\bar{L},i}^{(n)}=0,\qquad \omega_{P,i}^{(n)}=0
\nonumber
\eea
will provide constraints
$$
y_i^{(n)}=\frac{(-1)^n}{n!}\frac{d^n y_i^{(0)}}{dt^n},\qquad y_i^{(n)}=x_i^{(n)},\,\psi_i^{(n)},\,\bar{\psi}_i^{(n)},\,z_i^{(n)},\eqno{(A2)}
$$
as well as the equations of motion for the dynamical variables $x_i\equiv x_i^{(0)}$, $\psi_i\equiv \psi_i^{(0)}$, $\bar{\psi}_i\equiv \bar{\psi}_i^{(0)}$, $z_i\equiv z_i^{(0)}$,
\bea
\frac{d^{2l+1}x_i}{dt^{2l+1}}=0,\qquad\frac{d^{2l}\psi_i}{dt^{2l}}=0,\qquad \frac{d^{2l}\bar{\psi}_i}{dt^{2l}}=0,\qquad \frac{d^{2l-2\gamma+1}z_i}{dt^{2l-2\gamma+1}}=0.
\nonumber
\eea

The action functional which corresponds to these equations has the form
\bea
S=\frac{1}{2}\int\,dt\,\lambda_{ij}\,\left(x_i\frac{d^{2l+1}x_j}{dt^{2l+1}}-i\psi_i\frac{d^{2l}\bar{\psi}_j}{dt^{2l}}- i\bar{\psi}_i\frac{d^{2l}\psi_j}{dt^{2l}}+(-1)^{\gamma}\,z_i\frac{d^{2l-2\gamma+1}z_j}{dt^{2l-2\gamma+1}}\right).
\nonumber
\eea
This can be obtained from the one-form $\omega_M$ by taking into account (A2). In accord with (A1), this action is invariant under the transformations
\bea
\d t=a,\quad \d x_i=\sum_{n=0}^{2l}\tilde{a}_i^{(n)}t^n,\quad \d \bar{\psi}_i=\sum_{n=0}^{2l-1}\tilde{\bar{\xi}}_i^{(n)}t^n,\quad \d \psi_i=\sum_{n=0}^{2l-1}\tilde{\xi}_i^{(n)}t^n,\quad \d z_i=\sum_{n=0}^{2l-2\gamma}\tilde{b}_i^{(n)}t^n,
\nonumber
\eea
where $\tilde{\alpha}_i^{(n)}=(-1)^n \alpha_i^{(n)}$ and $\alpha_i^{(n)}=\{a_i^{(n)},\,\xi_i^{(n)},\,\bar{\xi}_i^{(n)},\,b_i^{(n)}$\}.

\end{document}